# PHOTOMETRY OF COMET C/2011 L4 (PANSTARRS) AT 4.4 - 4.2 AU HELIOCENTRIC DISTANCES.

Oleksandra Ivanova[a*], Serhii Borysenko[a], Alex Golovin[a]

[a] Main Astronomical Observatory of NAS of Ukraine, Akademika Zabolotnoho 27, Kyiv, 03680, Ukraine

[*] Corresponding Author. E-mail address: sandra@mao.kiev.ua

**Editorial correspondence to:**

Dr. Oleksandra Ivanova

Main Astronomical Observatory of NAS of Ukraine

Akademika Zabolotnoho str. 27

Kyiv, 03680

Ukraine

Phone: +380 44 526-47-69

E-Mail address: sandra@mao.kiev.ua

**ABSTRACT**

We present an analysis of the photometric data of comet C/2011 L4 (PANSTARRS) observed at heliocentric distance of 4.4 - 4.2 AU. The comet C/2011 L4 (PANSTARRS) shows one significant activity, despite of its quite large heliocentric distance. The color indexes, dust mass-loss rates and radius of the comet are measured.

## 1. Introduction

The Oort cloud and the Edgeworth-Kuiper belt are the two main reservoirs which supply the inner region of the Solar System with comets. These objects are considered to be relatively unmodified remnants of the early formation stages of the Solar System. The observation of long-period and dynamically new comets can provide information on the chemical composition of the cometary progenitors and the physical conditions of their origin. Being scattered into the inner region of our planetary system, some of the cometary nuclei become considerably active at heliocentric distances much larger than 4 AU. We have relatively small number of observed data on distant activity of long-period comets. But advances in technology development of modern detectors and using of large telescopes led to the discovery of many new objects beyond the orbit of Neptune, and a significant expansion of the list of comets that exhibit significant activity outside the Jupiter orbit. Using several methods of study of long-period comets activity at heliocentric distance from 3 to 26 AU information about dust characteristic, colors, morphology of the dust coma, estimate of radius of the nucleus (Meech, 1988, 1990, 2009; A'Hearn et al., 1995; Fulle et al., 1998; Rauer et al., 2003; Szabo et al., 2001,2002; Weiler et al., 2003; Korsun et al., 2003, 2006, 2010; Tozzi et al., 2003; Kawakita et al., 2004; Kelly et al., 2006; Mazzotta Epifani et al., 2009; Langland-Shula and Graeme H. Smith, 2011) were obtained. In this paper we present photometric observations of the Oort cloud comet C/2011 L4 (PANSSTARS) and the analysis of properties of the cometary dusty coma at the large heliocentric distance.

The comet C/2011 L4 (PANSTARRS) was discovered with Pan-STARRS 1 telescope on Haleakala, Maui, on 2011, June, 6. It was detected as a non-stellar object. Next observations of the object were obtained with Canada-France-Hawaii Telescope (Mauna Kea) on June 7.44 UT (Williams, 2011a). It was shown that this is the object with visible coma and faint tail. The detected object had 19.4$^{th}$ magnitude in R filter and it was located at the distance 7.17 and 8.16 AU from the Sun and the Earth, respectively. Williams (2011b) published the orbit of the comet on 2011 June 8 (two days after its discovery) based on 34 positions during the period from 2011 May 24 to June 8. This is a long period comet with inclination $i$= 84.2°, perihelion distance of $q$= 0.302 AU and e=1.0000318 (Nakano, 2013). The comet has perihelion on 10 March, 2013 and it passes very close to the Sun. It is very interesting object due to the high level of activity at the large heliocentric distance in different wavelengths.

## 2. Observations and reduction

The observations of comet C/2011 L4 (PANSTARRS) were made with the 2-m Faulkes Telescope South of Siding Spring Observatory (Australia) on May 31, 2012, when the heliocentric and geocentric distances of the comet were 4.46 and 3.45 AU, respectively. The next observations of the comet were made on June 21, 2011, when distance from the Sun and the Earth were 4.22 AU and 3.32 AU, respectively. The phase angle of the comet was 1.7° on 31 May and 7.3° on 21 June, respectively. The field was centered at the comet and 20 frames (with 60-sec exposures for each) were made.

A Fairchild CCD (486 BIDB) was used as a detector. The dimension of the image was 2048Ч2048 pixels and the scale was 0.3″/pix. The full field of view of the CCD is 10.0′×10.0′. The photometric data of the comet C/2011 L4 (PANSTARRS) was obtained in Bessel V and R broadband filters. To increase the signal/noise ratio of the observed data binning of 2×2 was applied to the photometric images. Standard bias subtraction and flat field reduction for all data were performed. All the frames with the images of the comet and standard stars were corrected for the zero-point and pixel sensitivity inhomogeneity using the master frames. Since the comet covered a small fraction of the field, we used routines of the IDL library (Goddard Space Flight Center) to calculate the sky background count (Landsman, 1993). After reduction, interframe median filtering was applied to the set of the images. This procedure allows us to increase the signal/noise ratio and to remove the field stars. The general view of the comet is presented in Fig.1.

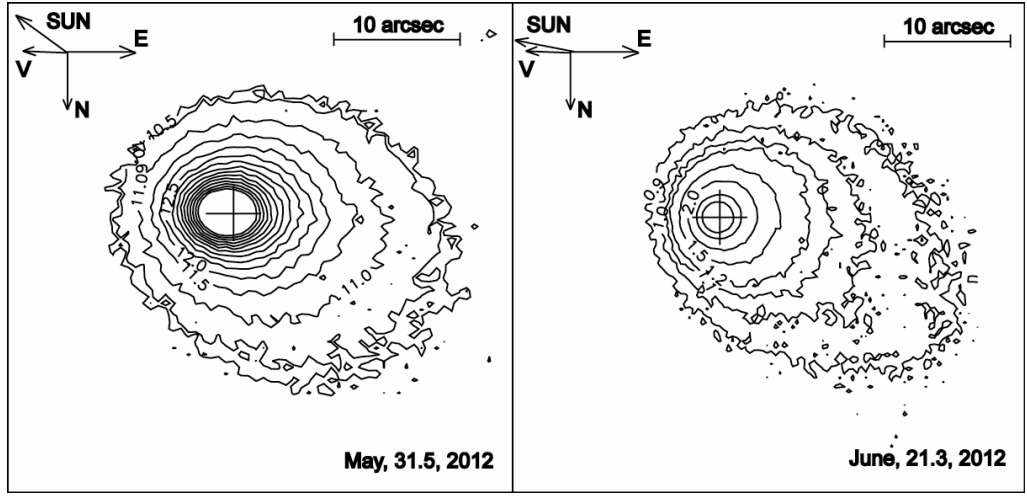

**Fig.1** Isophote contours of an image of comet C/2011 L4 (PANSTARRS) observed through a broad band R filter (for two observing sets). Isophots levels are in DN/sec. Celestial north, east, the motion V and sunward directions are denoted.

To perform an absolute flux calibration of the comet images three field stars were used for each observation set. The nights were photometric. For the aperture photometry of stars we used a diaphragm with the radius of 4″ due to the fact that the seeing value, measured as the average FWHM of several sample stars was about 1.9″ for 31 May 2012 and 1.7″ for 21 June 2012, respectively. The residual sky background was estimated with the use of an annular aperture. We used the USNO-A2.0 catalog for the photometric reference. Although this catalog is not photometric, there are several examples of its successful application for the standardization of photometric measurements of comets in the literature (Monet, 1998; Kidger, 2002; 2003; Hicks et al., 2007).

## 3. Observational results

We used the images obtained with the broadband V and R filters to measure the magnitude, effective radius of the nucleus, dust production rate of the comet C/2011 L4 (PANSTARRS). The cometary magnitude is given by

$$m_c = -2.5 \cdot \lg \left[ \frac{I_c(\lambda)}{I_s(\lambda)} \right] + m_{st} - 2.5 \cdot \lg P(\lambda) \cdot \Delta M \quad (1),$$

where $m_{st}$ is the magnitude of the standard star, $I_s$ and $I_c$ are the measured fluxes of the star and the comet in counts, respectively, $P$ is the sky transparency that depends on the wavelength, $\Delta M$ is the difference between the cometary and the stellar airmasses. As we used the field stars for calibration, the sky transparency is not considered. Using equation (1), the magnitude of the comet was calculated for an aperture radius of 9″. We obtained $m_V$ = 13.21±0.65, $m_R$ = 12.48±0.65 for observation on the 31st of May, and $m_V$ = 13.3±0.65, $m_R$ = 12.61±0.65 for observation on the 21st of June, respectively. Radii of the inner and outer rings are chosen large enough in order to avoid the possible contribution of the cometary coma in the measurement of the residual sky background around the image of the comet.

The colors of the cometary dust are redder than the Sun $V-R = 0.73^m \pm 0.91$ (May, 31) and $V-R = 0.69^m \pm 0.91$ (June, 21) for a 9″ photometric aperture. The Sun's color indexes $V-R = +0.52$ (Allen, 1976, Lamy et al., 1988) the Johnson filter system and $V-R = +0.35$ (Holmberg et al., 2006) in the Johnson-Cousins system.

The broadband V-filter covers strong emission bands due to the $C_2$ molecule in comets, so this filter is not used usually for analysis of dust color comet, with except distant comets. Our observations were made when the comet has heliocentric distance more than 4 AU. But some comets show $C_2$ emission, for example comet C/1995 Hale-Bopp (Rauer et al., 1997) and CO+, CN beyond 4 AU from the Sun (Cochran et al., 1980, 1982, 1991; Cook et al., 2005; Larson, 1980; Korsun, 2008). Unfortunately we don't have comet filters for observation the comet at Faulkes Telescope of South Siding Spring Observatory (Australia). C/2011 L4 (PANSTARRS) is new comet, and we don't have any information about it spectra too. But our estimate show that color indexes of the comet are redder than the Solar ones and we assume that the $C_2$ emission is not detected or is has small intensity and practically doesn't influence on the dust color the comet.

To compare the measurements of the dust productivity taken at different epochs, observing sites, and with different viewing geometries the parameter $Af\rho$ [cm] is used usually where $A$ is the average grain albedo, $f$ the filling factor in the aperture field of view, and $c$ the linear radius of the aperture (A'Hearn and Schleicher,1984). It is determined by the ratio of the effective cross-section of all grains, captured by the field of view of the detector to the projection of its field of view on the celestial sphere. The $Af\rho$ can be derived from the calculated photometric dust coma magnitude (Mazzotta Epifani et al., 2010):

$$Af\rho = \frac{4 \cdot r^2 \cdot \Delta^2 \cdot 10^{0.4 \cdot (m_{SUN} - m_R)}}{\rho} \qquad (2),$$

where $A$ is the average grain albedo, the filling factor $f$ is the total cross section of the grains in the field of view, $\rho$ is the radius of the field of view in cm, $r$ in AU and $\Delta$ in cm are the heliocentric and geocentric distances, respectively.

The eq. (2) is widely used for estimation $Af\rho$, when the dust production rate and the velocities of the ejected particles are constant (i.e. for the simple steady-state coma model). For such ideal steady-state coma, $Af\rho$ is aperture independent and can be used to derive the lower limit of the dust production rate (Bauer et al., 2003) and to compare activities of different comets. We calculate the $Af\rho$ values for a set of aperture values from 3S to 15S that used for images in the $R$ broadband filter (Fig.2) which isolate dust continuum of the comet.

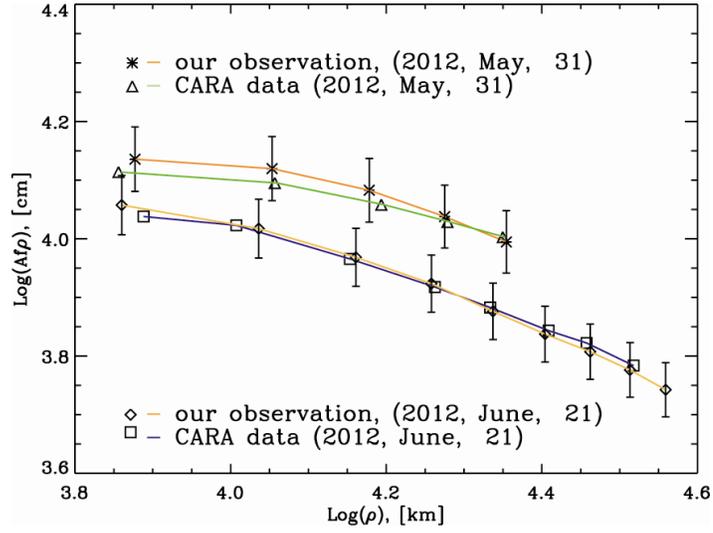

**Fig.2.** $Log-\log$ dependence of the $Af\rho$ parameter (in the R filter) on the aperture radius for comet C/2011 L4 (PANSTARRS) projected at the comets distances. Triangle and square mark the $Af\rho$ values for the comet provided by CARA team data.

We also obtained the radius of the nucleus $R_N$ (upper limits on it) basing on the observed magnitude $m_R$ and using (Russell, 1916; Jewitt, 2009):

$$p_r C = \frac{2.25 \cdot 10^{22} \cdot \pi \cdot r^2 \cdot \Delta^2 \cdot 10^{-0.4(m_d - m_{SUN})}}{10^{-0.4 \cdot \alpha \cdot \beta}} \quad (3),$$

where $p_r$ is the geometric grain albedo, $C$ is the geometrical cross-section of the nucleus in m², $r$ and $\Delta$ are the heliocentric and geocentric distances in AU, $m_{SUN}$ = −27.26 is the Bessel R-band magnitude of the Sun, $m_d = -2.5\lg_{10}\left(10^{-0.4m_2} - 10^{-0.4m_2}\right)$ the apparent R magnitude of the cometary coma in the annulus between apertures $\rho_1$ =3″ and $\rho_2$ =4.5″, $\alpha$ is the phase angle in degrees, and $\beta$ is the linear phase coefficient in magnitudes per degree. Equation (3) is often used for estimation of the effective radius of the cometary nuclei where $C = \pi \cdot R_N^2$. Usually this equation is used to obtain the effective radius of comets at large heliocentric distances which are not active or show low activity. Although the comet is rather active comet at heliocentric distance more than 4 AU, the upper limit of the cometary radius can be estimated. We calculate the radius of the nucleus using the geometric albedo and linear phase coefficient. First we used classical value of geometric albedo for nucleus $p_r$= 0.04, which is consistent with a similar study by Meech et al. (2009) for very distant comets. This value is close to the measured albedo values for other long period comets such as C/1983 H1 (0.03±0.01, Lamy et al., 2004), C/2001 OG108 (0.03 ± 0.005, Abell et al., 2003), and C/1985 O1 (0.04 ± 0.03, Jorda et al., 2000). For the phase coefficient we used the value 0.045 mag/°, which is the average of the measured phase coefficient values (Lamy et al. 2004).

Our calculation show that the values of apparent magnitude of the comet (Jewitt, 2009) are $m_d$ =14.24±0.65 for 31st of May 2012 and $m_d$ =14.65±0.65 magnitude for 21st of June, 2012; these values correspond to the upper limit of the cometary radius is $R_N$ = 59.6 km for observation on 31st of May 2012 (for $p_r$ =0.04 and $\beta$ =0.045), with the apparent R magnitude of the cometary coma in the annulus between apertures $\rho_1=3''$ and $\rho_2=4.5''$, and to $R_N = 50.7$ km for observation on 21st of June, 2012, respectively.

We estimated the dust productivity of the comet using the equation proposed by Jewitt (2009). According to the author, the effective cross-section and the total mass of dust particles is described by:

$$M_d = \frac{4}{3} \cdot \rho_{dust} \cdot (a_{min} \cdot a_{max})^{0.5} \cdot C \qquad (4)$$

where $\rho_{dust}$ is the grain density that does not depend on radius of the dust grain and is equal to 1000 kg/m³, $a_{min}$ and $a_{max}$ is the minimal and maximum radius of the grain particle which we adopt to be 0.1 μm and 1 cm (Grьn et al., 2001). To estimate the dust production, we introduce the grain residence time in the annulus defined by $\rho_1$ and $\rho_2$ are just:

$$\tau(r) = 1.5 \times \frac{10^{11} \cdot \Delta \cdot (\rho_2 - \rho_1)}{\upsilon(r)} \qquad (5)$$

where $\Delta$ is geocentric distance in AU, $\upsilon(r)$ -the radial outflow velocity of the dust grains from the nucleus, $\rho_1$ and $\rho_2$ are apertures of radii in radians (we used for calculation apertures $\rho_1=3''$ and $\rho_2=4.5''$). We used two values of radial outflow velocity of dust grains. The first value $\upsilon(r)=466$ m/s provides a simplified estimation of (an upper limit to) the dust-grain flow speed (Mazzota Epifani et al., 2010), based on measurements refer to gas for the comet Hale-Bopp (Biever et al., 2002). And the second value is $\upsilon(r)=50$ m/s. This value was obtained on basis of Probstein's theory (Probstein 1969; Fulle et al. 1998), which predicts that dust grains probably move more slowly than gas and than dust velocity close to 10% of the gas value. The ratio $dM_d/\tau(r)$ allows us to evaluate the dust production rate of the cometary nucleus (Table 1).

**Table 1. Model Mass-Loss Rates of the comet C/2011 L4 (PANSTARRS)**

| Data, UT | r, AU | Δ, AU | Ph[a], deg | $m_d$[b] | C, m² | τ, s | $M_d$, kg | $\frac{dM_d}{dt}$, kg/s |
|---|---|---|---|---|---|---|---|---|
| May, 31, 2012 | 4.457 | 3.450 | 1.7 | 14.24±0.65 | 1.12·10¹⁰ | 7.54·10⁴,c | 4.39·10⁸ | 5.82·10³,c |
|  |  |  |  |  |  | 8.1·10³,d |  | 5.42·10⁴,d |
| June, 21, 2012 | 4.222 | 3.324 | 7.3 | 14.65±0.65 | 8.06·10⁹ | 7.26·10⁴,c | 3.22·10⁸ | 4.43·10³,c |
|  |  |  |  |  |  | 7.8·10³,d |  | 4.13·10⁴,d |

[a] Phase angle
[b] the apparent magnitude of the cometary coma in the annulus between apertures $\rho_1=3''$ and $\rho_1=4.5''$ ($p_r=0.04$ and $\beta=0.045$)
[c] The dust-grain flow speed υ=50 m/s
[d] The dust-grain flow speed υ=466 m/s

## 4. Discussion and Conclusions

We present photometric observations of the Oort-cloud comet C/2011 L4 (PANSTARRS) obtained in May-June 2012. The R band images of the comet show a compact asymmetric dust coma on 31st of May, 2012 and asymmetric dust coma with small tail on 21st of June, 2012 (Fig.1).

Our estimations of the relative dust production rate of the comet show high activity level of the comet at heliocentric distant beyond 4 AU as well. The calculated $\log(Af\rho) - \log(\rho)$ relation is plotted in Fig. 2 for the projected distances from about 7000 km to 36000 km from the nucleus. To compare $Af\rho$ parameters derived from the different sites, the results from CARA project data base (Cometary Archives for Amateur Astronomers, http://www.uai.it) are also presented in the Fig. 2. Our results are very close to CARA data for observation period presented in the paper. Our results for the comet C/2011 L4 (PANSTARRS) support the idea that long-period comet are more active along the entire orbit and especially at large heliocentric distance (Meech 1988). For example, $Af\rho$ =1893 cm for C/2000 SV74 (LINEAR) observed at 4.04AU (Szabo et al., 2002), and $Af\rho$ = 3000 cm for C/1995 O1 (Hale-Bopp) observed at 4 AU (Weiler et al., 2003). Some results obtained for long-period comets at heliocentric distance above 5 AU show high level of $Af\rho$ too (Mazzotta Epifani et al., 2008; 2009; Meech et al., 2009, Tozzi et al., 2003).

To estimate the dust-loss rate of the comet C/2011 L4 (PANSTARRS), we used the first-order photometric model (Jewitt, 2009). Using the equation (4) for model calculations we obtained the upper limit of the dust mass loss-rate of the comet (see Tab.1). The high value of dust mass loss-rate close to our results were obtained also for long-period comets and some Centaurs (Fulle et al., 1998; Mazzotta Epifani et al., 2008; 2010; Jewitt, 2009), but this observation were carried for larger heliocentric distance, than our observation of the comet C/2011 L4 (PANSTARRS).

For cometary observations dominated by coma, we stress that the nuclear size estimates are upper limits ones. We obtained the upper limit of the radius of the comet C/2011 L4 (PANSTARRS) based on the photometric measurements $R_N$ <59.6 km for observation on 31st of May 2012 and $R_N$ <50.7 km for observation on 21st of June, 2012).

## Acknowledgments

The observations, made with the 2-m Faulkes Telescope of South Siding Spring Observatory (Australia), could only be performed with the support of the the Faulkes Telescope Project. We thank the CARA project members - A. Milani and G. Sostero for providing $Af\rho$ measurements for the comet. And we are grateful our Reviewers for the thoughtful reviews, whose comments and suggestions greatly aided this paper.

## 5. Reference

1. Abell, P. A., Fernandez, Y. R., Pravec, P., et al. 2003. Physical Characteristics of Asteroid-like Comet Nucleus C/2001 OG108 (LONEOS). Lunar and Planetary Science XXXIV. Abstract 1253 (Houston: Lunar and Planetary Institute).
2. A'Hearn, M. F., Schleicher, D. G., Millis, R. L., Feldman, P. D., Thompson, D. T., 1984. Comet Bowell 1980b. Astron. J. 89, 579–591.
3. A'Hearn M.F., Millis, R.L., Schleicher, D.G., Osip, D.J., Birch, P.V, 1995. The ensemble properties of comets: Results from narrowband photometry of 85 comets, 1976-1992. Icarus. 118, 223-270.
4. Allen, C.W., 1976. Astrophysical Quantities. Athlone Press, London, 689.

5. Bauer, J.M., Fernandez, Y.R., Meech, K.J., 2003. An optical survey of the active centaur C/NEAT (2001 T4). Astron. Soc. Pacific 115 (810), 981–989.

6. Biver, N., Воскелйе-Morvan, D., Colom, P., et al.2002. The 1995 2002 Long-Term Monitoring of Comet C/1995 O1 (HALE BOPP) at Radio Wavelength. Earth Moon and Planet, 90, 5-14.

7. Cochran, A.L., Barker, E.S., Cochran, W.D., 1980. Spectrophotometric observations of 29P/Schwassmann–Wachmann 1 during outburst. Astron. J. 85, 474–477.

8. Cochran, A.L., Cochran, W.D., Barker, E.S., 1982. Spectrophotometry of Comet Schwassmann–Wachmann 1. II – Its color and CO+ emission. Astrophys. J. 254, 816–822.

9. Cochran, A.L., Cochran, W.D., Barker, E.S., Storrs, A.D., 1991. The development of the CO+ coma of Comet P/Schwassmann–Wachmann 1. Icarus 92, 179–183.

10. Cook, J.C., Desch, S.J., Wyckoff, S., 2005. Visible and near infrared spectra of Comet 29P/Schwassmann–Wachmann 1. Bull. Am. Astron. Soc. 37, 645 (abstract).

11. Fulle, M., Cremonese, G., Bohm, C. 1998. The Preperihelion Dust Environment of C/1995 O1 Hale-Bopp from 13 to 4 A. Astron. J. 116, 1470-1477.

12. Grьn E., Hanner M. S., Peschke S. B., et al., 2001. Broadband infrared photometry of comet Hale-Bopp with ISOPHOT. Astron. Astrophys. 377, 1098-1118.

13. Hicks M. D., Bambery R. J., Lawrence K., J., et al., 2007. Near-nucleus photometry of comets using archived NEAT data. Icarus. 188, 457-467.

14. Holmberg J., Flynn C., Portinari L. 2006. The colours of the Sun. Monthly Notices of the Royal Astronomical Society. 367, 449-453.

15. Jewitt, D. C. 2009. The active centaurs. Astrophys. J. 137, 4296–4312.

16. Jorda, L., Lamy, P., Groussin, O., et al. 2000. ISOCAM Observations of Cometary Nuclei. Proceedings of ISO Beyond Point Sources: Studies of Extended Infrared Emission, ed. R. J. Laureijs et al.,ESA-SP, 455.

17. Kawakita, H., Watanabe, J., Ootsubo, T., et al. 2004. Evidence of Icy Grains in Comet C/2002 T7 (LINEAR) at 3.52 AU. Astrophys. J, 601, L191-L194.

18. Kelley, M. S., Woodward, C. E., Harker, D. E., et al. 2006. A Spitzer Study of Comets 2P/Encke, 67P/Churyumov-Gerasimenko, and C/2001 HT50 (LINEAR-NEAT).Astrophys. J, 651, 1256-1271.

19. Kidger, M. R. 2002. Spanish Monitoring of Comets: Making Sense of Amateur Photometric Data Earth, Moon and Planets. 90, 259-268.

20. Kidger. M. R. 2003. Dust production and coma morphology of 67P/Churyumov-Gerasimenko during the 2002-2003 apparition. Astron. and Astrophys. 408, 767-774.

21. Korsun, P. P., Chorny, G. F. 2003. Dust tail of the distant comet C/1999 J2 (Skiff). Astron. And Astrophys. 410, 1029-1037.

22. Korsun, P. P., Ivanova, O. V., & Afanasiev, V. L. 2006. Cometary activity of distant object C/2002 VQ94 (LINEAR). Astron. Astrophys. 459, 977-980.

23. Korsun, P.P., Ivanova, O.V., Afanasiev, V.L., 2008. C/2002 VQ94 (LINEAR) and 29P/Schwassmann–Wachmann 1 – CO+ and N2+ rich comets. Icarus. 198, 465–471.

24. Korsun, P.P., Kulyk, I.V., Ivanova, O.V., Afanasiev, V.L., Kugel, F., Rinner, C.,Ivashchenko, Yu.N., 2010. Dust tail of the active distant Comet C/2003 WT42 (LINEAR) studied with photometric and spectroscopic observations. Icarus. 210, 916–929.

25. Lamy, P. L., H. Pedersen, and R. Vio. 1988. The dust tail of comet P/Halley in April 1986. Exploration of Halley's Comet. Springer Berlin Heidelberg. 661-664.

26. Lamy, P. L., Toth, I., Fernandez, Y. R., & Weaver, H. A. 2004. The sizes, shapes, albedos, and colors of cometary nuclei. Comets II, M. C. Festou, H. U. Keller, and H. A. Weaver (eds.), University of Arizona Press, Tucson, 745 pp., 223-264.

27. Landsman, W.B. 1993. The IDL Astronomy User's Library. Astron. Data Analysis Software and Systems II, A.S.P. Conference Series, R. J. Hanisch, R. J. V. Brissenden, and Jeannette Barnes, eds. 52, 246-248.

28. Langland-Shula L. E, Smith, G. H. 2011. Comet classification with new methods for gas and dust spectroscopy. Icarus. 213,Iss. 1, 280-322.

29. Larson, S., 1980. CO+ in Comet Schwassmann–Wachmann 1 near minimum brightness. Astrophys. J. 238L, L47–L48.

30. Mazzotta Epifani, E., Palumbo, P., Capria, M. T., et al., 2008. The distant activity of Short Period Comets– II. MNRAS. 390, 1, 265.

31. Mazzotta Epifani, E., Palumbo, P., Capria, M. T., Cremonese, G., Fulle, M., Colangeli, L., 2009. The distant activity of the Long Period Comets C/2003 O1 (LINEAR) and C/2004 K1 (Catalina). Astron. Asrophys. 502, 355–365.

32. Mazzotta Epifani E., Dall'Ora M., di Fabrizio L., et al., 2010. The Activity of Comet C/2007 D1 (LINEAR) at 9.7 AU from the Sun. Astron. Astrophys. 513, A33.

33. Meech, K. J. 1988. Light Curves of Periodic vs Dynamically New Comets as a Function of R. BAAS, 20, 835.

34. Meech, K. J. 1990. Activity on Comet Cernis (1983 XII) at Large Distances. BAAS, 22, 1103.

35. Meech, K.J., Pittichova, J., Bar-Nun, A., Notesco, J., Laufer, D., Hainaut, O.R., Lowry, S.C., Yeomans, D.K., Pitts, M., 2009. Activity of comets at large heliocentric distances pre-perihelion. Icarus 201, 719–739.

36. Monet. D. G. 1998. The 526,280,881 Objects In The USNO-A2.0 Catalog. Bull. Amer. Astronom. Soc. 30, 1427.

37. Nakano,S. 2013. http://www.oaa.gr.jp/~oaacs/nk/nk2386.htm

38. Probstein, R. F. 1969. in Problems of Hydrodynamics and Continuum Mechanics, ed. F. Bisshopp et al. (Philadelphia: Soc. Indust. Appl. Math.). 568

39. Rauer, H., Biver, N., Crovisier, J., et al., 1997. Millimetric and optical observations of Chiron. Planet. Space Sci., 45, 799-805.

40. Rauer, H., Helbert, J., Arpigny, C., et al. 2003. Long-term optical spectrophotometric monitoring of comet C/1995 O1 (Hale-Bopp). Astron. Astrophys., 397, 1109-1122.

41. Russel H. N., 1916. On the Albedo of the Planets and their Satellites. Astrophys. J. 43, 173-196.

42. Szabo, G. M., Csak, B., Sarneczky, K., Kiss, L. L. 2001. Photometric observations of distant active comets. Astron. Astrophys., 374, 712-718.

43. Szabo, G. M., Kiss, L. L., Sarneczky, K., Sziladi, K. 2002. Spectrophotometry and structural analysis of 5 comets. Astron. Astrophys., 384, 702-710.

44. Tozzi, G. P., Boehnhardt, H., Lo Curto, G. 2003. Imaging and spectroscopy of comet C/2001 Q4 (NEAT) at 8.6 AU from the Sun. Astron. Astrophys., 398, L41-L44.

45. Weiler M., Rauer H., Knollenberg J., et al., 2003. The dust activity of comet C/1995 O1 (Halle-Bopp) between 3AU and 13AU from the Sun. Astron. Astrophys. 403, 313-322.

46. Williams, G.V., 2011a. Comet P/2011 L4 (PANSTARRS). MPEC. 2011-L33.

47. Williams, G.V., 2011b. Observations and orbits of comets. MPEC. 2011-J16.